\documentclass[sigconf,screen,nonacm]{acmart}

\usepackage[para]{footmisc}
\usepackage{enumitem}
\usepackage{algorithm}
\newcommand{\xiw}[1]{#1}
\newcommand{\sm}[1]{#1}

\usepackage{algpseudocode}
\usepackage{float}
\newfloat{algorithm}{t}{lop}

\usepackage{todonotes}
\usepackage{booktabs}
\usepackage{balance}
\usepackage{siunitx}
\usepackage{amsmath}
\usepackage{balance}

\usepackage{tabularx,arydshln}
 \usepackage{array,multirow,graphicx}

\newcommand{\add}[1]{#1}

\AtBeginDocument{%
  \providecommand\BibTeX{{%
    \normalfont B\kern-0.5em{\scshape i\kern-0.25em b}\kern-0.8em\TeX}}}

\setcopyright{acmcopyright}
\copyrightyear{2018}
\acmYear{2018}
\acmDOI{XXXXXXX.XXXXXXX}

\acmJournal{JACM}
\acmVolume{37}
\acmNumber{4}
\acmArticle{111}
\acmMonth{8}

\acmPrice{15.00}
\acmISBN{978-1-4503-XXXX-X/18/06}

\begin{document}

\title{Online Distillation for Pseudo-Relevance Feedback}

\author{Sean MacAvaney}
\authornote{Both authors contributed equally to this research.}
\email{sean.macavaney@glasgow.ac.uk}
\orcid{0000-0002-8914-2659}
\affiliation{
\institution{University of Glasgow}
\country{United Kingdom}
}
\author{Xi Wang}
\authornotemark[1]
\orcid{0000-0001-5936-9919}
\email{xi-wang@ucl.ac.uk}
\affiliation{
\institution{University College London}
\country{United Kingdom}
}

\renewcommand{\shortauthors}{MacAvaney and Wang}

\begin{abstract}
Model distillation has emerged as a prominent technique to improve neural search models. To date, distillation taken an \textit{offline} approach, wherein a new neural model is trained to predict relevance scores between arbitrary queries and documents. In this paper, we explore a departure from this offline distillation strategy by investigating whether a model for a specific query can be effectively distilled from neural re-ranking results (i.e., distilling in an \textit{online} setting). Indeed, we find that a lexical model distilled online can reasonably replicate the re-ranking of a neural model. More importantly, these models can be used as \textit{queries} that execute efficiently on indexes. This second retrieval stage can enrich the pool of documents for re-ranking by identifying documents that were missed in the first retrieval stage. Empirically, we show that this approach performs favourably when compared with established pseudo relevance feedback techniques, dense retrieval methods, and sparse-dense ensemble ``hybrid'' approaches.
\end{abstract}


\maketitle

\section{Introduction}

\looseness -1 Search result ranking is a critical component of information retrieval systems, and recent advancements in neural networks, especially pre-trained language models, have shown great promise in improving its effectiveness~\cite{khattab2020colbert, choi2021improving}. Despite their potential, optimal ranking outcomes often require extensive model training, particularly for large-scale parameter models. Re-rankers, which reorder a set of documents (i.e., pseudo-relevant documents) retrieved by another component, can be a cost-effective solution~\cite{macavaney2020efficient, wang2011cascade}. However, one significant drawback of re-rankers is that they completely discard documents that were not identified in the first stage of retrieval. To overcome this recall problem, various strategies have been proposed, including dense retrieval (e.g.,~\cite{Xiong2020ApproximateNN}), learned sparse retrieval (e.g.,~\cite{10.1145/3404835.3463098}), and document re-writing (e.g.,~\cite{Nogueira2019DocumentEB}). Nonetheless, these approaches require considerable computation and storage overheads, which can be particularly burdensome for large document collections.

\begin{figure}
\centering
\includegraphics[scale=0.35]{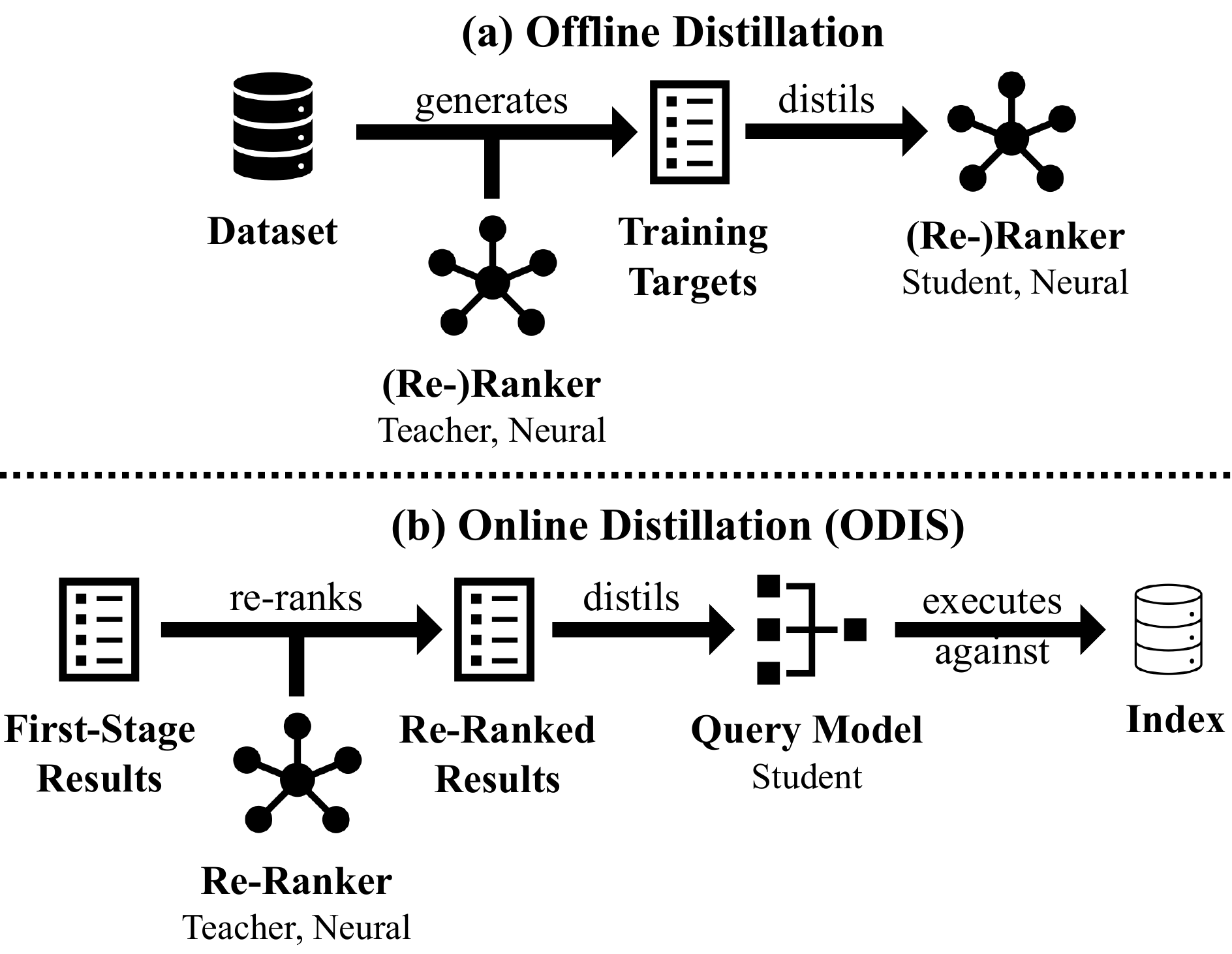}
\caption{Comparison between (a) Offline Distillation and (b) Online DIStillation (\xiw{ODIS}, ours).}
\label{fig:high_level}
\end{figure}

Meanwhile, there has been considerable attention paid to model distillation for ranking. This involves training a smaller neural model, referred to as the ``student'' model, to mimic the behaviour of a larger neural ``teacher'' model~\cite{sebastian2020improving}. The primary focus of the student model is to emulate the teacher model's generalization capabilities. As a result, we often observe that the student models perform just as well, if not better, than their teacher models~\cite{Hinton15Distilling}. However, existing efforts in this area have primarily been conducted offline, meaning that the distillation occurs over a large set of queries from a training dataset. This can introduce additional costs to deploy the distilled student model. Further, the costs are paid each time a new teacher model is trained, e.g., for updates to match recent search trends. Figure~\ref{fig:high_level}(a) provides a graphical overview of the offline distillation process. It shows that two separate stages are required: the first optimizes the heavy teacher model to achieve promising results, and the second conducts the distillation to obtain a lighter student model.

\looseness -1 In this work, we bridge the use of pseudo-relevance feedback and model distillation by adding a new twist to both settings: using model distillation to provide pseudo-relevance feedback. Instead of ranking all documents for each query using an expensive model, we distil a lightweight single-query model. This model approximates the ranking scores for pseudo-relevant documents and generates a new ranked list \sm{over} the entire document corpus \sm{for the specific query}. This direction has several advantages: it is efficient enough to train during query processing itself, and with a suitable distilling process, the resulting student model itself is able to be executed \textit{as a query} over the index. In contrast with existing pseudo-relevance feedback approaches, this setting allows for optimisation over arbitrary \sm{relevance} functions, \sm{rather than relying on top-$k$ heuristics}. Indeed, in this work, we \sm{demonstrate that distilling a student model comprised of a linear combination of term weights} via minimising the rank-biased score differences \add{can be effective}. \sm{We find this setting} outperforms the traditional pseudo relevance feedback technique of assuming \sm{a} fixed top-$k$ documents are relevant. Figure~\ref{fig:high_level}(b) shows an overview of our approach, Online DIStillation (ODIS). \add{Specifically, ODIS distils a lightweight model of the current result set at query-time, rather than prior work that has explored the distillation of a larger model prior to indexing.}

In summary, our formulation of distillation differs from past efforts in several significant ways: (1) the distillation occurs online rather than offline, (2) the distillation targets improvement in recall for a specific query rather than building a model capable of scoring a variety of queries, and (3) the student model itself can be efficiently executed over an index.

We experimentally demonstrate the effectiveness of our Online Distillation (ODIS) approach. We observe that ODIS has a strong ability to approximate various leading retrieval pipelines in ranking relevant documents. Moreover, by applying the learned lightweight scoring function (i.e., distilled student model) to rank the entire corpus efficiently, we observe consistent and significant performance improvements, especially with respect to the recall of missing relevant documents. Overall, our results highlight the potential of ODIS as a valuable technique for improving retrieval effectiveness.\footnote{\add{During peer review of this work, a contemporaneous preprint~\cite{DBLP:journals/corr/abs-2305-11744} was published that proposed a similar strategy to ours.}}

\section{Related Work}

In this section, we provide a relevant context from prior work for our online distillation approach, ODIS, and elaborate on past works on Pseudo-Relevance Feedback (PRF) and Model Distillation (MD). 

\begin{table}
\centering
\caption{Table of Symbols.}
\label{tab:symbols}
\begin{tabular}{cl}
\toprule
Symbol & Description \\
\midrule
$\mathbf{Q}$ & Query \\
$\mathbf{D}$ & Document \\
$\mathbf{C}$ & Corpus of documents \\
$\mathbf{C}^x$ & A subset of $\mathbf{C}$, filtered and scored by function $x$ \\
\midrule
$\mathcal{S}$ & Scoring function (e.g., a cross-encoder) \\
$\mathcal{R}$ & Retrieval function (e.g., BM25) \\
$\mathcal{O}$ & Distilled scoring model \\
$x\text{>>}y$ & $x$ re-ranked by $y$ \\
\midrule
$\theta$ & Model parameters of $\mathcal{O}$ \\
$\mathcal{L}$ & Loss function for distilling $\mathcal{O}$ \\
$w_{\mathbf{D}_1, \mathbf{D}_2}$ & Weight of document pair $\mathbf{D}_1$ and $\mathbf{D}_2$ \\
$r$ & L1 regularisation rate for distilling $\mathcal{O}$ \\
\bottomrule
\end{tabular}
\end{table}

\paragraph{Pseudo Relevance Feedback}

PRF techniques are commonly employed to enhance search results by assuming the relevance of the top-ranked documents in the initial search results to the user's query. The existing PRF methods can be categorised as per their used types of feedback, learning strategies and relevance criteria of documents. For instance, Bo1~\cite{amati2002probabilistic} and RM3~\cite{nasreen2004umass} leveraged the statistical information of pseudo-relevant documents for query expansion, which also serves as a strong baseline (BM25 with RM3) for newly introduced retrievers~\cite{wang2021pseudo,chandradevan2022learning}. With the increasing trend of neural techniques, many neural PRF approaches are emerging, such as CEQE~\cite{naseri2021ceqe}, which leverages a language model and pseudo-relevant documents for query expansion through contextualised semantic similarity. Other similar techniques are NPRF~\cite{li2018nprf}, BERT-QE~\cite{zheng2020bert}, ColBERT-PRF~\cite{wang2021pseudo} and ANCE-PRF~\cite{li2022improving}, but they differ in applying neural techniques for updating pseudo-relevance documents~\cite{zheng2020bert}, identifying meaningful expansions~\cite{wang2021pseudo} or updating representations~\cite{li2022improving} for a given query. Despite their success, the major limitations of these neural approaches are their time and memory inefficiency~\cite{zhan2022learning}. As a result, it is challenging to apply these techniques to evaluate the relevance of all documents for each query and instead, they tend to focus only on pseudo-relevant documents, which may not guarantee the retrieval of the most related documents for a given query. In contrast, our proposed method, ODIS, distils and approximates the relevance scores between queries and the top-ranked documents and extends the computation to all documents for an overall evaluation so as to address the above concern. 

The most similar prior work to ours is \citet{Lin2021TheST}, which proposed training linear classifiers to re-rank search results at runtime. Our work differs substantially from this, both in motivation and function. First, \sm{ODIS} is designed to function as a query over an inverted index (i.e., re-write the query), rather than simply re-rank results. Special considerations were made for this -- particularly by controlling the sparsity of the model. Further, rather than a classification approach, which assumes a certain number of documents are relevant and non-relevant, we take a distillation approach, wherein the model training objective aims to mimic the complete ranking.

\paragraph{Model Distillation (MD)} MD is a popular method that operates in a teacher-student mode, where a large neural model (teacher) transfers its learned knowledge to a smaller, more efficient model (student)~\cite{Hinton15Distilling}. Due to its efficiency and complexity-saving nature, it has been widely used in addressing multiple tasks~\cite{zhou2019understanding,fu2021lrc,park2019relational}, including our focus on document retrieval~\cite{thibault2021splade,sebastian2020improving,lin2020distilling}.
For example, TCT~\cite{lin2020distilling} distils the learned knowledge from the ColBERT model~\cite{khattab2020colbert} to generate a similar relevance distribution -- judged by KL divergence -- on documents for queries. Unlike TCT,~\citet{choi2021improving} encouraged the student model to generate similar representation embeddings for the `[CLS]' token, queries and documents for the ad hoc retrieval. In contrast with existing approaches that train student models capable of handing a variety of queries, ODIS distils single-query student models, aimed at improving the recall for individual queries.

\begin{algorithm}
\caption{Online Distillation}\label{alg:main}
\begin{algorithmic}
\Require
  $\mathbf{Q}$ query,
  $\mathbf{C}$ corpus,
  $\mathcal{S}$ scoring function,
  $\mathcal{R}$ retrieval function
\State $\mathbf{C}^\mathcal{R} \gets$ top results from $\mathbf{C}$ for $\mathbf{Q}$ using $\mathcal{R}$
\State $\mathbf{C}^{\mathcal{R}\text{>>}\mathcal{S}} \gets$ score $\mathbf{C}^\mathcal{R}$ for $\mathbf{Q}$ using $\mathcal{S}$
\State $\mathcal{O} \gets$ distil $\mathbf{C}^{\mathcal{R}\text{>>}\mathcal{S}}$
\State $\mathbf{C}^\mathcal{O} \gets$ top results from $\mathbf{C}$ for $\mathbf{Q}$ using $\mathcal{O}$
\State $\mathbf{C}^{\mathcal{O}\text{>>}\mathcal{S}} \gets$ score $\mathbf{C}^\mathcal{O}$ for $\mathbf{Q}$ using $\mathcal{S}$
\State $\triangleright$ Note: documents in $\mathbf{C}^\mathcal{O}\cap\mathbf{C}^\mathcal{R}$ need only be scored once by $\mathcal{S}$
\State $\triangleright$ Note: the total number of documents scored by $\mathcal{S}$ can be capped
\Ensure $\mathbf{C}^{\mathcal{R}\text{>>}\mathcal{S}} \cup \mathbf{C}^{\mathcal{O}\text{>>}\mathcal{S}}$
\end{algorithmic}
\end{algorithm}

\section{ODIS: Online Distillation}

Consider a scoring function $\mathcal{S}:\textbf{Q}\times\textbf{D}\mapsto \mathbb{R}$, which maps query $\textbf{Q}$ and document $\textbf{D}$ to a real-valued relevance score (notation in Table~\ref{tab:symbols}). When $\mathcal{S}$ is expensive to run (such as a cross encoder model), it is impractical to score all documents in corpus $\textbf{C}$ for a given query. Therefore, a subset of the corpus $\mathbf{C}^\mathcal{R} \subset \textbf{C}$ is selected for scoring, using the top results of efficient retrieval $\mathcal{R}:\textbf{Q}\times\textbf{D}\mapsto \mathbb{R}$ (e.g., BM25). This ``re-ranking'' or ``telescoping'' approach is the predominant approach for using expensive models like $\mathcal{S}$ for ranking.

However, $\mathbf{C}^\mathcal{R}$ will not necessarily contain all the documents that $\mathcal{S}$ would have scored the highest across all of $\mathbf{C}$. Online Distillation (ODIS) aims to better approximate the full ranking results of $\mathcal{S}$ over $\textbf{C}$ for a given $\textbf{Q}$ by \textit{distilling} a lightweight, single-query model $\mathcal{O}:\mathbf{D}\mapsto\mathbb{R}$. 
For a well-designed $\mathcal{O}$, we can efficiently execute it over the entire corpus to produce a new ranked list $\mathbf{C}^\mathcal{O} \subset \mathbf{C}$. Since $\mathcal{O}$ is distilled to mimic the rankings of $\mathcal{S}$, we can expect $\mathbf{C}^\mathcal{O}$ to contain new documents that are potentially relevant to $\mathcal{S}$. In other words, we expect $\mathbf{C}^\mathcal{O}\setminus\mathbf{C}^\mathcal{R}$ to contain documents that $\mathcal{S}$ will score highly. Given this, we can then score $\mathbf{C}^\mathcal{O}\cup\mathbf{C}^\mathcal{R}$ using $\mathcal{S}$ as the final output, in a typical PRF fashion. An overview of this process is given in Algorithm~\ref{alg:main}, and we detail the distillation procedure in the following section.

\subsection{Distillation Procedure}

$\mathcal{O}$ can be any learnable function, but for efficiency, we consider it as a linear combination of features, parameterised by $\theta$. The goal of the distillation process is to find a set of parameters that minimise the loss $\mathcal{L}$ between the document scores produced by $\mathcal{S}$ and those for $\mathcal{O}$: $\min_{\theta}
\mathcal{L}\big(\mathcal{S}, \mathbf{Q}, \mathbf{C}^\mathcal{R}\big)$. An approximate solution can be found using stochastic gradient descent approaches, such as Adam~\cite{kingma2014adam}.

Finding a linear combination of field values that reconstructs the (often arbitrary) scale of scores from $\mathcal{S}$ could be challenging, depending on the nature of the fields. We therefore propose using a weighted pairwise preference loss function:
\begin{equation}
\mathcal{L}\big(\mathcal{S}, \mathbf{Q}, \mathbf{C}^\mathcal{R}\big)
=
\sum^{\mathbf{C}^\mathcal{R} \times \mathbf{C}^\mathcal{R}}_{(\mathbf{D}_1, \mathbf{D}_2)}
w_{\mathbf{D}_1, \mathbf{D}_2} (\mathcal{O}(\mathbf{D}_2) - \mathcal{O}(\mathbf{D}_1))
\end{equation}
where $w$ is a weighting score for the pair of documents. For \add{$w$}, we use the difference in reciprocal ranks (from $\mathcal{S}$) between the two documents, which prioritises pairs with at least one document scored near the top of the ranked list.

In our experiments, we use token-level TF-IDF features,\footnote{ODIS could be applied in other settings as well, such as over dense document embedding features to find a strong query vector for dense retrieval; we leave this exploration to future work.} allowing the parameters $\theta$ to define the weights of query terms that can be executed against a lexical index for efficient retrieval. We introduce two practical modifications to ODIS to account for this setting. First, since most retrieval engines limit token weights to be positive, we include a ReLU activation function over $\theta$, which clips negative weights up to 0. Second, since queries with many tokens are costly to execute, we add a $\ell_1$ regularisation component to the loss, which pushes parameter weights towards zero. The weight of this component with respect to the data loss is controlled by hyperparameter $r$. We default $r=1$, but automatically increase $r$ by a factor of 10 if the model converges without achieving a target level of sparsity $t$, the maximum number of non-zero token weights. \add{Put another way, if the model converges without the desired sparsity, the model will continue to train, but with a higher regularisation rate, encouraging a model that is more sparse the next time it converges.}

\section{Experimental Setup}

\looseness -1 We run experiments to answer the following questions about ODIS:

\begin{enumerate}
\item[RQ1] \textbf{Can models distilled in an online fashion mimic the rankings from relevance models?} The replication of the rankings is likely necessary for models to perform well as queries, and the answer to this question is non-obvious, considering the complexity of neural relevance models.
\item[RQ2] \textbf{Do online distilled models generalise enough to identify new relevant documents from the corpus?} This question is at the core of our method, since to goal is to improve the recall of re-ranking pool.
\item[RQ3] \textbf{What is the computational and storage overheads of the approach?} Overheads need to be low enough to allow for practical use.
\end{enumerate}

\textbf{Datasets and Measures.} We test our approach on the TREC Deep Learning 2019 and 2020 (DL19\footnote{\texttt{msmarco-passage/trec-dl-2019}} and DL20\footnote{\texttt{msmarco-passage/trec-dl-2020}}) passage ranking datasets~\cite{trec-dl19,trec-dl20}. We report nDCG and Recall (with a minimum relevance score of 2) cut off at 1000 results.\footnote{The official measure for the datasets are nDCG cut off at 10, but we observed virtually no difference among systems at this depth, so we explore deeper ranking quality.} We also measure the Rank Biased Overlap (RBO, $p=0.99$)~\cite{Webber2010ASM} between given result sets and \sm{exhaustive rankings of the scorer over the entire corpus to check how well the approach approximates the complete rankings}.

\textbf{Pipelines.} We test various pipelines that re-rank a cross-encoder model (CE\footnote{\texttt{cross-encoder/ms-marco-MiniLM-L-6-v2}}), which have shown promise as a teacher model with rich interactions between transformer-encoded queries and documents~\cite{lin2021batch}. As initial result sets, we test: a lexical model (BM25), two dense models (BE, a bi-encoder\footnote{\texttt{sentence-transformers/msmarco-distilbert-base-v2}}, and BE$_\text{CE}$, a bi-encoder distilled from the CE\footnote{\texttt{sentence-transformers/msmarco-distilbert-dot-v5}}), and an ensemble sparse-dense hybrid system of BM25 and BE (via reciprocal rank fusion at $k=60$). For context, we also include the results of CE after an exhaustive search over the entire corpus.

\textbf{Baselines.} We compare ODIS with several competitive general-purpose pseudo-relevance feedback methods. RM3~\cite{nasreen2004umass} and Bo1~\cite{amati2002probabilistic} are lexical PRF methods that identify and weight salient terms from the top documents, as re-ranked by CE. Graph Adaptive Reranking (GAR, \cite{mao2021generation}) is a \xiw{recent} PRF approach incorporated as a step that pulls in documents nearby the top results during the re-ranking process. \xiw{Several other neural PRF techniques, such as ANCE-PRF~\cite{yu2021improving} and ColBERT-PRF~\cite{wang2021pseudo}, have been proposed in the literature. \sm{However, these approaches are not general --- they are tied to a particular classes of models (single-representation or multi-representation dense retrieval, respectively.) We, therefore, also focus on only general-purpose PRF methods, including} GAR~\cite{mao2021generation}, which represents \sm{a strong,} recent advance. We plan to extend ODIS to include dense features in future work and compare it with neural techniques as the next stage of this work.}

In all pipelines, we score a maximum of 1000 documents using CE. To validate the effectiveness of various PRF approaches, for RM3, Bo1, and ODIS, we consume up to half the budget (500 documents) on the first-stage results and the remaining budget on the documents retrieved during PRF. Similarly, we use the ``alternate'' variant of GAR, which consumes half the budget from the initial retrieval and half from nearby documents\sm{, in an iterative fashion}.

\textbf{Parameters and Tuning.} ODIS was developed on the BM25 pipeline over DL19 using a different cross-encoder model \add{(MonoT5~\cite{DBLP:journals/corr/abs-2003-06713})}, while the DL20 dataset and other pipelines were held out for \sm{final} evaluation only. \add{This experimental setup was developed to ensure that neither the method itself nor the hyper-parameters are over-fit to our evaluation data; we demonstrate that the method and settings transfer zero-shot to other pipelines and datasets.} To facilitate a fair comparison among methods, we equally tuned the number of feedback terms for RM3, Bo1, and ODIS on the BM25 pipeline over DL19. We observed stable and strong effectiveness for all three models at 50 feedback terms, which lead to a consistent application of this setting across all pipelines. As is common for PRF, we also tuned the $\lambda$ parameter for all applicable methods (including ODIS), which represents the weight of the new query terms WRT the original query terms.

\section{Results and Analysis}

\begin{table}
\centering
\caption{Distillation quality of ODIS over various retrieval pipelines on DL19. Each row compares the pipeline to one with ODIS, either re-ranking the original results or using the ODIS model to retrieve from the corpus. * indicates significant $\Delta$nDCG values \add{(Student's paired t-test, $p<0.05$)}.}
\label{tab:rq1}
\begin{tabular}{lrrrrr}
\toprule
& \multicolumn{2}{c}{Re-Rank} & \multicolumn{3}{c}{Retrieve} \\
\cmidrule(lr){2-3}\cmidrule(lr){4-6}
Pipeline & RBO & $\Delta$nDCG & RBO & Overlap & +Rel/q \\
\midrule
BM25 >> CE & 0.584 & *$-$0.017 & 0.377 & 0.269 & 10.2 \\
BE >> CE & 0.623 & $-$0.006 & 0.385 & 0.197 & 15.7 \\
BE$_{\text{CE}}$ >> CE & 0.593 & $-$0.008 & 0.415 & 0.250 & 9.7 \\
BM25+BE >> CE & 0.571 & $-$0.006 & 0.407 & 0.300 & 7.3 \\
\bottomrule
\end{tabular}
\end{table}

We begin by testing whether an ODIS lexical model is capable of adequately \textit{fitting the cross-encoder model} \textbf{(RQ1)}. To test this, we train ODIS models on the re-ranked results from the cross-encoder model and then compare the effectiveness of the model as per its ranking results to that of the original ones. Table~\ref{tab:rq1} presents the results, both in re-raking and retrieval settings. When re-ranking, the ODIS model achieves RBO scores of between 0.571 and 0.623, when compared with the cross encoder's ranking -- a reasonably strong overlap.\footnote{For reference, two rankings that are identical aside from the items at rank 1 and 12 swapped have an RBO of 0.6.} Not all differences matter in terms of ranking result quality, however. We therefore also measure the difference in nDCG score ($\Delta$nDCG) between the original and ODIS results, and find that ODIS only degrades the quality by up to 0.017 nDCG. In three of the four cases, the differences are not statistically significant. These results answer \textbf{RQ1}: the rankings of a cross-encoder can be successfully distilled down into a linear combination of up to 50 TF-IDF-weighted terms. This conclusion is remarkable, given the complexity of cross-encoder models \sm{and the simplicity of the distilled model}. \add{Still, to achieve the highest-quality results in the final ranking, it's the teacher model should be used.}

\begin{table}
\centering\small
\caption{Effectiveness of ODIS and baselines over various pipelines. Significant differences between ODIS and corresponding baselines are indicated as superscripts:  CE$^c$, RM3$^r$, Bo1$^b$, GAR$^g$ (\add{Student's} paired t-test, $p<0.05$). RBO compares each combined result list with the exhaustive CE search.}
\label{tab:main}
\scalebox{0.94}{
\begin{tabular}{clrrrrrr}
\toprule
&& \multicolumn{2}{c}{DL19} & \multicolumn{2}{c}{DL20} \\
\cmidrule(lr){3-4}\cmidrule(lr){5-6}
\multicolumn{2}{l}{System} & nDCG & R@1k & nDCG & R@1k & RBO \\
\midrule

\parbox[t]{1mm}{\multirow{6}{*}{\rotatebox[origin=c]{90}{Lexical}}} & BM25 & 0.602 & 0.755 & 0.596 & 0.805 & 0.347 \\
&\hspace{4pt} >> CE & 0.703 & 0.755 & 0.717 & 0.805 & 0.711 \\
&\hspace{4pt} >> RM3$_{\text{CE}}$ & 0.759 & 0.845 & 0.770 & 0.887 & 0.761 \\
&\hspace{4pt} >> Bo1$_{\text{CE}}$ & 0.745 & 0.815 & 0.757 & 0.857 & 0.749 \\
&\hspace{4pt} >> GAR$_{\text{CE}}$ & 0.753 & 0.839 & 0.757 & 0.878 &\bf  0.772 \\
&\hspace{4pt} >> ODIS$_{\text{CE}}$ &\bf  $^{cbg}$0.768 &\bf  $^{cb}$0.859 &\bf  $^{cbg}$0.785 &\bf  $^{cb}$0.909 & $^{cb}$0.769 \\
\midrule
\parbox[t]{1mm}{\multirow{6}{*}{\rotatebox[origin=c]{90}{Dense}}} & BE & 0.607 & 0.734 & 0.594 & 0.773 & 0.467 \\
&\hspace{4pt} >> CE & 0.674 & 0.734 & 0.679 & 0.773 & 0.822 \\
&\hspace{4pt} >> RM3$_{\text{CE}}$ & 0.768 & 0.882 &\bf  0.781 &\bf  0.925 & 0.876 \\
&\hspace{4pt} >> Bo1$_{\text{CE}}$ & 0.770 & 0.877 & 0.772 & 0.899 & 0.885 \\
&\hspace{4pt} >> GAR$_{\text{CE}}$ & 0.746 & 0.849 & 0.740 & 0.867 & 0.867 \\
&\hspace{4pt} >> ODIS$_{\text{CE}}$ &\bf  $^{cg}$0.777 &\bf  $^{cg}$0.894 & $^{cbg}$0.779 & $^{cg}$0.911 &\bf  $^{crg}$0.889 \\
\midrule
\parbox[t]{1mm}{\multirow{6}{*}{\rotatebox[origin=c]{90}{Dense (Distilled)}}} & BE$_{\text{CE}}$ & 0.698 & 0.818 & 0.696 & 0.843 & 0.632 \\
&\hspace{4pt} >> CE & 0.728 & 0.818 & 0.728 & 0.843 & 0.916 \\
&\hspace{4pt} >> RM3$_{\text{CE}}$ & 0.777 & 0.904 &\bf  0.784 &\bf  0.930 & 0.918 \\
&\hspace{4pt} >> Bo1$_{\text{CE}}$ & 0.777 & 0.897 & 0.775 & 0.907 & 0.922 \\
&\hspace{4pt} >> GAR$_{\text{CE}}$ & 0.758 & 0.880 & 0.754 & 0.884 & 0.918 \\
&\hspace{4pt} >> ODIS$_{\text{CE}}$ &\bf  $^{cg}$0.783 &\bf  $^{c}$0.909 & $^{cg}$0.781 & $^{cg}$0.917 &\bf  $^{crbg}$0.924 \\
\midrule
\parbox[t]{1mm}{\multirow{6}{*}{\rotatebox[origin=c]{90}{Ensemble}}} & BM25+BE & 0.725 & 0.856 & 0.697 & 0.871 & 0.534 \\
&\hspace{4pt} >> CE & 0.755 & 0.856 & 0.752 & 0.871 & 0.886 \\
&\hspace{4pt} >> RM3$_{\text{CE}}$ & 0.776 & 0.887 & 0.784 & 0.921 & 0.869 \\
&\hspace{4pt} >> Bo1$_{\text{CE}}$ & 0.769 & 0.873 & 0.774 & 0.898 & 0.867 \\
&\hspace{4pt} >> GAR$_{\text{CE}}$ & 0.769 & 0.881 & 0.767 & 0.902 &\bf  0.889 \\
&\hspace{4pt} >> ODIS$_{\text{CE}}$ &\bf  $^{cb}$0.780 &\bf  $^{cb}$0.894 &\bf  $^{cbg}$0.790 &\bf  $^{cb}$0.927 & $^{crbg}$0.875 \\
\midrule
\multicolumn{2}{l}{CE (Exhaustive)} & 0.768 & 0.894 & 0.765 & 0.906 & 1.000 \\

\bottomrule
\end{tabular}
}
\end{table}

However, the ODIS models might be overfitting to the CE rankings -- the models are not valuable unless they can \textit{identify new relevant documents from the corpus} \xiw{\textbf{(RQ2)}}. The retrieval results from Table~\ref{tab:rq1} show that the ODIS models indeed identify new potentially-relevant documents. The retrieval RBO is markedly lower than in the re-ranking setting, and the overlap ratio between the newly retrieved documents with the original results only reach up to 0.3\sm{---both of which demonstrate that new documents are introduced in the rankings}. Among such new documents retrieved, between 7.3 and 15.7 of them are relevant documents that were missed in the first stage. \add{These newly-retried documents yield a marked improvement in overall system recall; e.g., from 0.805 R@1000 to 0.909 R@1000 for the BM25 pipeline, as shown in Table~\ref{tab:main}.}

These newly retrieved relevant documents are not particularly helpful unless they can be successfully incorporated into the rankings by the cross-encoder model. \sm{To address} this concern, Table~\ref{tab:main} presents the overall effectiveness of the ODIS-augmented retrieval pipelines and baselines. We observe that ODIS significantly improves over the base pipeline in both benchmarks and all four pipelines. ODIS also performs favourably compared to relevant PRF baselines. RM3 is overall the strongest baseline in terms of retrieval effectiveness; indeed, there are never significant differences between ODIS and RM3 in terms of nDCG or R@1k. However, in three out of four pipelines, ODIS provides results that are significantly closer to the exhaustive CE rankings (RBO), meaning that it provides a more ``faithful'' approximation of the full results. Meanwhile, it provides a significant improvement over Bo1, another lexical PRF technique in both pipelines where lexical signals are already present in the original rankings (BM25 and BM25+BE), but not in pipelines where only dense signals are used for the initial ranking. Finally, ODIS typically outperforms GAR in terms of nDCG (7 of 8 tests), while not requiring pre-computed nearest neighbours.

To answer \textbf{RQ2}, online distilled models identify new relevant documents that enable effective rankings of the scoring model. 
Further, ODIS outperforms strong baseline PRF approaches in terms of ranking effectiveness (or produces results that are more faithful to the complete ranking when there is no significant difference in ranking effectiveness), making \xiw{ODIS} an attractive alternative to existing techniques.

Next, we consider \textit{the overheads of performing ODIS} \xiw{\textbf{(RQ3)}}. In terms of storage, ODIS incurs no extra overhead compared to existing lexical PRF approaches (RM3 and Bo1); they all make use of a direct index to look up the tokens that occur in documents and a inverted index to perform retrieval. The computational overhead can be broken down into training and retrieval stages. When we tested running ODIS on an NVIDIA 3090 GPU, we found that ODIS training \sm{only} takes 98ms per query on average, albeit with only 20\% GPU utilisation. Given that such pipelines involve a cross-encoder scorer, requiring a GPU for ODIS is reasonable, \add{and only represents a fraction of the time spent scoring documents}. However, RM3 and Bo1 are far less costly, requiring only 4ms on average to rewrite the query, without the need for a hardware accelerator. Future work could investigate techniques to reduce the computational overhead of ODIS distillation, or make better use of hardware acceleration to increase utilisation. On the other hand, both RM3 and Bo1 also involve a secondary retrieval stage, so no computational overhead with respect to baselines in incurred at that stage. Meanwhile, GAR incurs very low query-time overhead (around 20ms/query total), but requires an expensive offline nearest neighbour computation. In summary, to answer \textbf{RQ3}, ODIS adds computational overhead compared to baselines in the query rewriting process but does not add relative overhead in terms of storage or retrieval.

\section{Conclusions and Future Work}

We proposed an online model distillation approach, and showed how it can be effectively used for pseudo-relevance feedback. The effectiveness of ODIS is favourable compared to established approaches, by either producing more relevant results or (when relevance is comparable), producing rankings that are more faithful to an exhaustive cross encoder search. This work sets the stage for a variety of follow-up work to be conducted. For instance, future work could explore whether ODIS can effectively distil dense query vectors, or even non-linear student models. Future work could also explore whether ODIS models could themselves be used as training data for first-stage query rewriting.

\section*{Acknowledgements}

\add{We thank the anonymous reviewers for their helpful feedback on this manuscript.}
\balance
\bibliographystyle{ACM-Reference-Format}
\balance
\bibliography{sample-base}

\end{document}